# An Ontology Model for Organizing Information Resources Sharing on Personal Web

Istiadi[1], and Azhari SN[2]

[1]Department of Electrical Engineering, University of Widyagama Malang,
Jalan Borobudur 35, Malang 65128 Indonesia,
email: istiadi@widyagama.ac.id

[2] Department of Computer Science and Electronics, Gadjah Mada University, Yogyakarta,
Sekip Unit III FMIPA Gedung Selatan Sleman, Yogjakarta 55281 Indonesia,
email: arisn@ugm.ac.id

**Abstract.** *Retrieve information resources made by the machine processing may refer to multiple sources. A personal web as part of information resources in the Internet requires a feature that can be understood by computer machines. Therefore, in this paper an ontology semantic web approach is used to map the resources in a meaningful scheme. In the design of concept, resources on the web are viewed as documents that have some property and ownership. Domain interest or web scope is used to describe a classification of resources that navigate into relevant documents. If instances are completed to the concept, then the ontology file can be loaded and shared as annotation on personal web. This allows computer machine to query multiple ontology from different personal webs that use it.*

## 1. INTRODUCTION

Presence of the Internet has provided opportunities to share information which is generally widely publicized through the web. In addition, web sites developed by institutions or organizations are also mostly personal web sites that appear. This is because it is more easy to develop web through web development services for free or more affordable web development costs. In fact, many educational institutions provide space for their staffs to create personal webs. In this situation, the existence of personal webs will complement information needed by the public.

The orientation in the development of web today is not just for users (people) consumption but more that namely it furthermore processed by computer engine (software agent) [1]. This is related to increasingly complex information needs. So that some data openly available on web (open content) enables them to be furthermore processed or combined to obtain other forms or contexts of information. Therefore support on web needs to be completed in a formal semantic format (Semantic Web), thus providing space for the computer engine to access efficiently.

Semantic Web needs to be developed and implemented on a variety of web-based information providers, including the personal web. Implementation semantic web has been researched and developed mainly in the organizational service systems area , for example in e-learning system [4], digital library [2] [6] and so on. Personal web has many diverse potentials, besides personal information given, it may also be information from their knowledge and experience shared openly [7]. Aa wide variety of content may be complementary with each other on specific domain of knowledge or information. So that when supported by a certain processing it may form a kind of library from distributed information resources.

Because its personal website need to be supported by a scheme of organizing information resources that can be processed by computer machines.

In this paper the ontology model will be described to organize and to support information resource shared on the personal web which is a form of application of the semantic web. So that the personal resources on the web is expected to be accessible by computer machines for certain purposes such as providing a referral list of some the personal web content which will be exemplified in the application of the proposed model of ontology.

## 2. SEMANTIC WEB AS ENABLER ENGINE PROCESSING

The Use of Semantic Web may be enabled by a set of standards which are coordinated by the World Wide Web Consortium (W3C). The most important standards in building the Semantic Web are XML, XMLSchema, RDF, OWL, and SPARQL.

Extensible Markup Language (XML) is a markup language designed to be an easy facilities to send documents over the Web. However, the XML standard has no semantic constraints on the meaning of the document. An XML Schema is a language used to define a set of rules (schema) that must be followed by the XML document. The structure of XML documents must be made in accordance with the defined schema.

Resource Description Framework (RDF) is a specification created by W3C as a general method for modeling collection of information by using the syntax format. The basic idea of RDF is to make a statement on a Web resource in the form of expression "subject-predicate-object" or summarized by term SPO. In the RDF terminology, SPO is often referred to as N-triple. RDF Schema can be viewed as a data dictionary or vocabulary for describing properties and classes of RDF resources.

Web Ontology Language (OWL) is a language that can be used by applications which does not just display information in humans, but which need to process information content. Ontology itself can be defined as a way to describe concepts and relationships of domains that become a concern. It contains descriptions of classes, properties, and individuals [5]. This description can help the computer system use those terms with an easier way. By using OWL, it can add vocabulary, in addition to formal semantics that has been previously created using XML, RDF, and RDF Schema.

Given the semantic representation of the standard formats (RDF /RDF Schema and OWL), then the access to resources can be made by using a SPARQL query [3]. The machine will get data or more definitive information in a way that enables the implementation of query criteria in the form of a triple. SPARQL supports the use of union operation to allow merging data from multiple sources.

The Development of semantic web can be done in a simple way using a text editor but it requires attention more than using a tool. One of the tools for assisting the development of semantic web is a Protege. Protege provides complete support to use of standards that have been recommended by W3C so that users can develop and conduct testing such as by using SPARQL query [5].

In this paper the authors use the Protege for developing ontology, applying objects and data property. Ontology models that produced can be linked to the personal web as annotation so it is not necessary to change web content. To try query processing is used SPARQLer query engine that are available online for producing combined data from multiple ontology placed on some personal web host.

## 3. MODELING ONTOLOGY FOR PERSONAL WEB

Personal web is a web made by an individual who is more personal information than the information company, organization, or institution. Content on personal web varies but generally it includes aspects of information shared in a certain scope according to interest of their owners [11]. In general, the personal web space contains some documents, among which has a link (hyperlink) or references both internal and external links [10]. Each web has an identifier document called URI [9] to be accessed or referenced as a link.

Based on this description, there are some aspects that become a part of existence of a personal web. These aspects include the web entity itself, aspect of ownership of resources on the web, aspect of scope that describes classification of resources, and aspects of content which is a set of information resources that are viewed as documents. These aspects are described in the concept model as shown in the picture below.

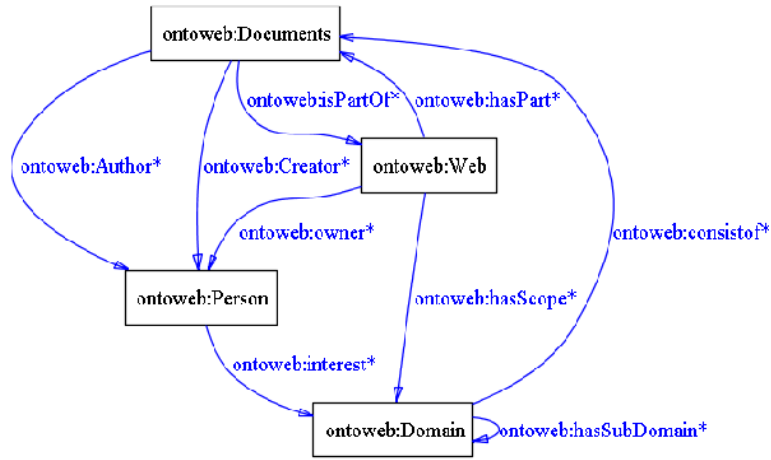

Figure 1. Concept Design of Ontology

Aspects of web entities was declared as a Class Web. These entities are viewed as a part of a large resource available on the Internet. Web entities may be identified with a URI and is also used as a base (base) URI for internal resources. Class Webs have a owners that relate to Person Class, and have a scope relationship with Class Domain. Property data type associated with Class Web is a title of web and a brief description of web. Below is shown a script of Class Web and its properties in the form of RDF / XML.

Listing 1. Defining Class Web as Part of Model

```
…
<owl:Class rdf:ID="Web"/>
<owl:ObjectProperty rdf:ID="owner">
    <rdfs:domain rdf:resource="#Web"/>
    <rdfs:range rdf:resource="#Person"/>
  </owl:ObjectProperty>
<owl:ObjectProperty rdf:ID="hasPart">
    <rdfs:domain rdf:resource="#Web"/>
    <owl:inverseOf rdf:resource="#isPartOf"/>
    <rdfs:range rdf:resource="#Documents"/>
  </owl:ObjectProperty>
```

```
<owl:ObjectProperty rdf:ID="hasScope">
    <rdfs:domain rdf:resource="#Web"/>
    <rdfs:range rdf:resource="#Domain"/>
</owl:ObjectProperty>
<owl:DatatypeProperty rdf:ID="webAbout">
    <rdfs:domain rdf:resource="#Web"/>
    <rdfs:range rdf:resource="&xsd;string"/>
</owl:DatatypeProperty>
<owl:DatatypeProperty rdf:ID="webTitle">
    <rdfs:domain rdf:resource="#Web"/>
    <rdfs:comment xml:lang="en">null</rdfs:comment>
    <rdfs:range rdf:resource="&xsd;string"/>
</owl:DatatypeProperty>
...
```

Aspect of ownership is represented by Class Person as the owner and the creator of web and resources. Class Person is an individual set of Object Names. Class Person also has interest relation to domain content areas covered on the web. Property data type is associated with Class Person of which is such a contact email and can be equipped with some other relevant properties of profile information. Below is shown script of Class Person and its properties in form of RDF / XML.

Listing 2. Defining Class Person as Part of Model

```
...
<owl:Class rdf:ID="Person"/>
    <owl:ObjectProperty rdf:ID="interest">
    <rdfs:domain rdf:resource="#Person"/>
    <rdfs:range rdf:resource="#Domain"/>
</owl:ObjectProperty>
<owl:ObjectProperty rdf:ID="Author">
    <rdfs:domain rdf:resource="#Documents"/>
    <rdfs:range rdf:resource="#Person"/>
</owl:ObjectProperty>
<owl:DatatypeProperty rdf:ID="prsMbox">
    <rdfs:domain rdf:resource="#Person"/>
    <rdfs:range rdf:resource="&xsd;string"/>
</owl:DatatypeProperty>
...
```

Aspects of scope given by Class Domain are areas of interest and cover parts of relevant documents. This is also a means for classifying contents that allows computer engine to directly refer to more relevant documents. Property data type associated with Class Domain is a brief description of Domain. Objects on class domain can have a subdomain relation to refer to a narrower scope. Below is shown a script of Class Domain and its properties in the form of RDF / XML.

Listing 3. Defining Class Domain as Part of Model

```
...
<owl:Class rdf:ID="Domain"/>
  <owl:ObjectProperty rdf:ID="consistof">
      <rdfs:domain rdf:resource="#Domain"/>
      <rdfs:range rdf:resource="#Documents"/>
  </owl:ObjectProperty>
  <owl:ObjectProperty rdf:ID="hasSubDomain">
      <rdfs:domain rdf:resource="#Domain"/>
      <rdfs:range rdf:resource="#Domain"/>
  </owl:ObjectProperty>
  <owl:DatatypeProperty rdf:ID="domDescription">
      <rdfs:domain rdf:resource="#Domain"/>
      <rdfs:range rdf:resource="&xsd;string"/>
  </owl:DatatypeProperty>
...
```

Aspect of content is related to information resources form of documents contained on the web. This aspect is represented by Class Document. It has relationship with creator or author and is a part of scope of a specific domain. Some properties associated with the Document Class is title, document type, description ,resources identifier (URI), a link to another resource, date, and publications. Below is shown script of Class Document and its properties in form of RDF / XML.

Listing 4. Defining Class Document as Part of Model

```
...
<owl:Class rdf:ID="Documents"/>
  <owl:DatatypeProperty rdf:ID="docURI">
      <rdfs:domain rdf:resource="#Documents"/>
      <rdfs:range rdf:resource="&xsd;string"/>
  </owl:DatatypeProperty>
<owl:ObjectProperty rdf:ID="isPartOf">
      <rdfs:domain rdf:resource="#Documents"/>
      <owl:inverseOf rdf:resource="#hasPart"/>
      <rdfs:range rdf:resource="#Web"/>
</owl:ObjectProperty>
<owl:ObjectProperty rdf:ID="Creator">
      <rdfs:domain rdf:resource="#Documents"/>
      <rdfs:range rdf:resource="#Person"/>
</owl:ObjectProperty>
<owl:DatatypeProperty rdf:ID="docDate">
      <rdfs:domain rdf:resource="#Documents"/>
      <rdfs:range rdf:resource="&xsd;date"/>
</owl:DatatypeProperty>
<owl:DatatypeProperty rdf:ID="docDescription">
      <rdfs:domain rdf:resource="#Documents"/>
      <rdfs:range rdf:resource="&xsd;string"/>
</owl:DatatypeProperty>
  <owl:DatatypeProperty rdf:ID="docLink">
      <rdfs:domain rdf:resource="#Documents"/>
      <rdfs:range rdf:resource="&xsd;string"/>
  </owl:DatatypeProperty>
  <owl:DatatypeProperty rdf:ID="docPublish">
```

```
    <rdfs:domain rdf:resource="#Documents"/>
    <rdfs:range rdf:resource="&xsd;string"/>
  </owl:DatatypeProperty>
  <owl:DatatypeProperty rdf:ID="docTitle">
    <rdfs:domain rdf:resource="#Documents"/>
    <rdfs:range rdf:resource="&xsd;string"/>
  </owl:DatatypeProperty>
  <owl:DatatypeProperty rdf:ID="docType">
    <rdfs:domain rdf:resource="#Documents"/>
    <rdfs:range rdf:resource="&xsd;string"/>
  </owl:DatatypeProperty>
…
```

Ontology  above is a conceptual model that provides some Class and relationships to accommodate associated instances. People can implement it by placing  objects and data of resources available on their web in accordance with its classes and properties.

## 4. RESULTS AND DISCUSSION

### 4.1 Case Example

To illustrate how the ontology model  is used, the following explanation of case ontology implementing  is given on query mechanism of multiple personal web ontology. For example, there are three owners of personal Web respective interests in the world of LinuxOS, but each has concentrated on different fields such as relating to use a different distro like Ubuntu, Debian, or IGOS. They have practical experiences related to the linux world, written in the form of the articles which are distributed through the web. For example, problem solving installing web server ,  applications, and hardware  are based using distro.

From those experiences, they can share information to complement each other as an alternative that can be referenced by users accessing to one of them. They can use the ontology of the proposed scheme by applying the information resources on their web  as its instance. Furthermore  the ontology is a OWL  file placed on each host (see Figure 2). Thus alternatives are available  for the processing machine to access resources on their website through its file.

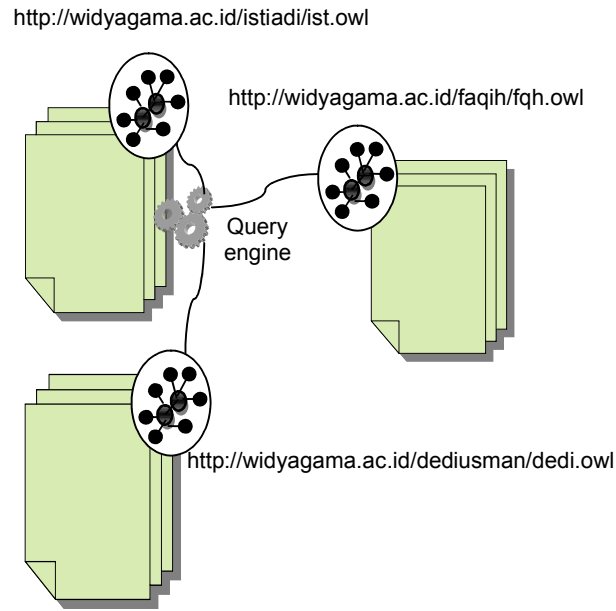

Figure 2.  Ontology models applied on some personal web  for processing by query engine

From those ontologies a new service on their web may be added. The service as like list of contents from their web can be developed based on distributed ontology. On their web it can be added query processing mechanism from multiple of ontology. Query is made to provide referrals based on a list of domain scope which can be forwarded  to a more specific scope based on sub domain . The sub domain scope  is used for directing to the relevant documents. Union operation from query of multiple ontologies  is used to produce  the list of content that combine the form s of their webs.

**4.2. Query Multiple Personal Web Ontology**

To obtain proof of how the query can be done as in the above scenario, The SPARQLer  available on-line (http://sparql.org/sparql.html) is used as query engine. SPARQLer provides a form to write the query statement and provides a choice of output formats such as JSON, XML, Text, CSV, and TSV. The following figure shows the query statement that is used in processing SPARQLer.

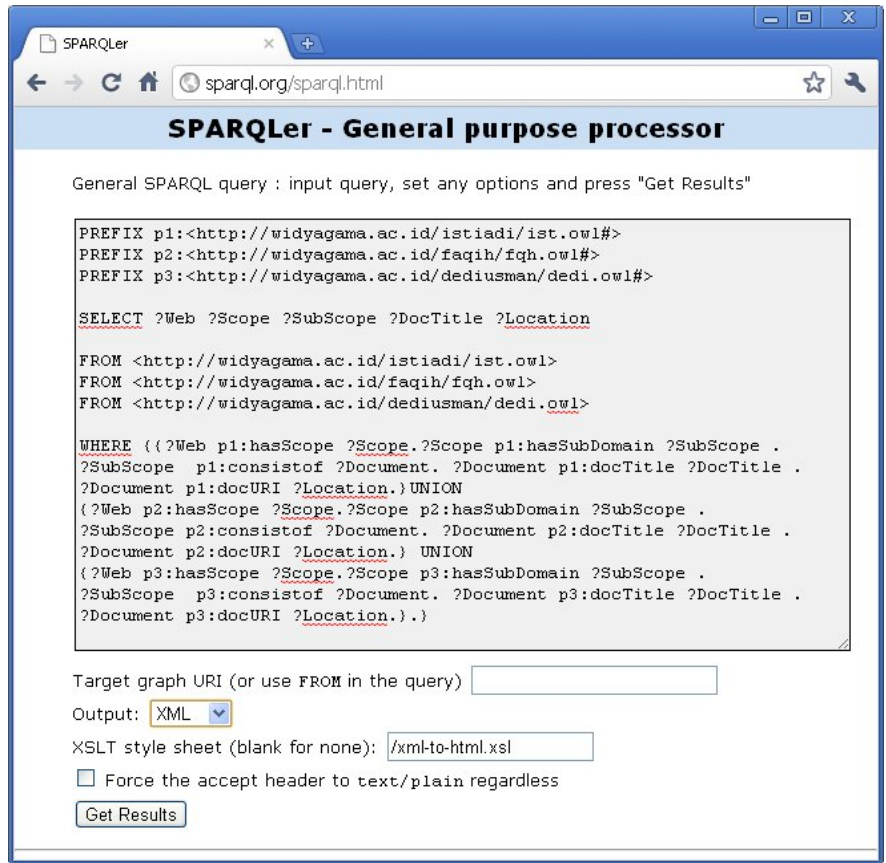

Figure 3. Query multiple ontology personal web using SPARQLer

Query statement shown above is to access three different entities of owl files (ist.owl, fqh.owl, dedi.owl). PREFIX statement is used to indicate an abbreviation URI name space of the owl files referred. SELECT statement is used to select a subset of data query results to be displayed, in this case referred to the selected web (indicated by ?Web), the scope of the web (indicated by ?Scope), a more specific scope (indicated by ?SubScope), title-related documents (indicated by ?DocTitle), and the location of the document (indicated by ?location). WHERE statement is used to determine the conditions of selection based on the query graph data pattern . UNION operation is used to combine queries from multiple different graph. The results of the query execution in a tabular format is shown below.

Table 1. Query Result

| Web | Scope | SubScope | DocTitle | Location |
|-----|-------|----------|----------|----------|
| WebofIstiadi | Linux_OS | Ubuntu | "Installing LAMP on Ubuntu 9.10" | "http://widyagama.ac.id/istiadi/installingLAMP_Ubuntu/" |
| WebofIstiadi | Linux_OS | Ubuntu | "Installing Printer driver C90" | "http://widyagama.ac.id/istiadai/Printerdriverc90/" |
| faqihweb | Linux_OS | Igos | "Membuat Web Server pada IGOS" | "http://widyagama.ac.id/faqih/igos_webserver/" |
| faqihweb | Linux_OS | Ubuntu | "Dependency problem installing | "http://widyagama.ac.id/faqih/dependency_webserver/" |

| | | | Web Server" | |
|---|---|---|---|---|
| Webofdedi | Linux_OS | Debian | "How to configure connection to Repository" | "http://widyagama.ac.id/dediusman/debian_repo/" |
| Webofdedi | Linux_OS | Debian | "How to Install web server @ Debian" | "http://widyagama.ac.id/dediusman/Install_webserver/" |

The results above is a combination of the three personal web information through query mechanism in the form of tables. The result describes the contents of some web resources that inform the document classification based on its scope. With that information, it might provide a better alternative for the user because it is a combination of several sources based on context of information. With the choice of a personal web it will be more complete because it it related to information provided by other web personal on the same of context.

Simple implementation of the use of these query may be made by utilizing SPARQLer engine with copying the URI that query generated as a link that can fitted on each of their personal web concerned. By accessing the link it will redirect the browser accessing to SPARQLer that will process it to generate a list of contents of their webs.

## 5. CONCLUSION

Organization of resource sharing on personal website to accessible by a machine is enabled with an ontology approach that provides a conceptual of resources and relationships. Personal resources on web were viewed as a collection of documents on a specific domain as part of web scope or personal interest. This domain acts as a navigator which directs computer machine to find relevant documents. Classification can be narrowly done by defining the subdomain. The simple application could be used to provide a list of the contents by combining some personal webs that apply ontology through a query mechanism.

In the future the ontology should be integrated from application of the web system, so that when the content or web document is changed then ontology will also be automatically updated. The use of ontology for supporting such an integration is exemplified, its application needs to use subject terms reference to the domain scope that is uniform or accommodates alternative that should be considered in the query statement.